# Experimental Study of Diamond Like Carbon (DLC) Coated Electrodes for Pulsed High Gradient Electron Gun

Martin Paraliev, Christopher Gough, Sladjana Ivkovic, Frederic Le Pimpec
Paul Scherrer Institute
5232 Villigen, Switzerland
martin.paraliev@psi.ch

*Abstract*-For the SwissFEL Free Electron Laser project at the Paul Scherrer Institute, a pulsed High Gradient (HG) electron gun was used to study low emittance electron sources. Different metals and surface treatments for the cathode and anode were studied for their HG suitability. Diamond Like Carbon (DLC) coatings are found to perform exceptionally well for vacuum gap insulation. A set of DLC coated electrodes with different coating parameters were tested for both vacuum breakdown and photo electron emission. Surface electric fields over 250MV/m (350 - 400kV, pulsed) were achieved without breakdown. From the same surface, it was possible to photo-emit an electron beam at gradients up to 150MV/m. The test setup and the experimental results are presented.

## I. INTRODUCTION

Within the SwissFEL Free Electron Laser project a pulsed electron gun was built and commissioned [1]. High electrical gradient is used to accelerate rapidly the electron bunch in order to preserve beam emittance. Electrode shaping is used to provide additional electrostatic beam focusing in the accelerating gap. Different electrode materials and surface treatments were measured for their High Gradient (HG) breakdown strength. Surface electric fields over 250MV/m were achieved with hydrogenated amorphous Diamond Like Carbon (DLC) coated electrodes.

DLC coated metal surface was used as a photocathode and beam was extracted at wide range of surface gradient, from 25 to 150MV/m.

Following successful results of an elliptical geometry, hollow electrode geometry was developed in order to study electron emission from different samples and Field Emitting Arrays (FEAs).

## II. TEST SETUP OVERVIEW

The test electron gun consists of 500kV High Voltage (HV) pulse generator, HG accelerating diode, laser illumination system, two-cell Radio Frequency (RF) accelerating cavity and electron beam diagnostics line. The diode vacuum chamber (Fig 1.) is designed to allow rapid exchange of the electrodes and it is used for both HG and electron beam experiments.

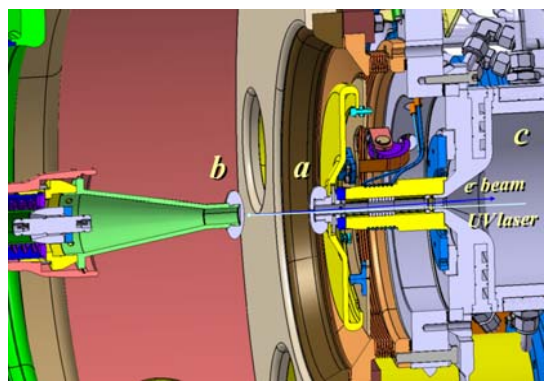

Figure 1. HG acceleration diode: *a* anode, *b* cathode, *c* two cell RF accelerating structures

The HV pulse generator provides a damped asymmetric oscillatory pulse that is applied to the cathode to establish HG in the accelerating gap. Depending on emitter type, a laser pulse or electrical signal is used to trigger the cathode emission on the crest of the first negative oscillation. Typical waveform of the accelerating voltage is shown on Fig. 2. The accelerating gradient is set using both the amplitude of the applied voltage pulse and anode-cathode distance.

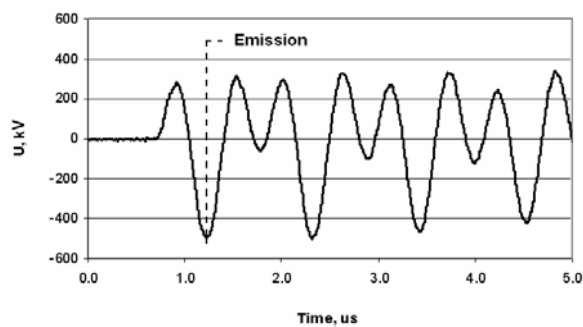

Figure 2. Typical acceleration voltage waveform.

On leaving the diode, the electron bunch enters the RF cavity for further acceleration. The electron beam is characterized by downstream diagnostics, including electron beam emittance, profile, energy, energy spread and charge measurements. A directional X-ray detector is used to monitor the bremsstrahlung from the dark current in the accelerating diode and in the RF cavity. Without the laser irradiation, the accelerating diode is used to make HG electrodes tests.

In order to compare breakdown strength of different materials and surface finishes a three phase test procedure was defined (Fig. 3.).

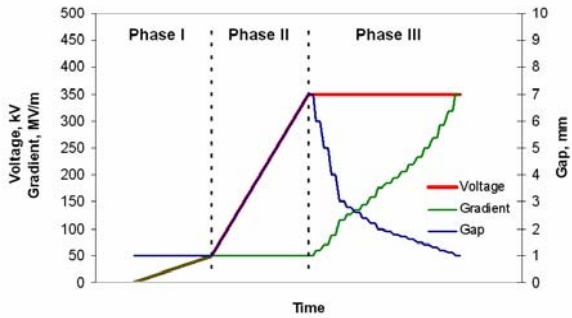

Figure 3.  HG test procedure with three phases.

Phase I – anode-cathode gap is set to 1mm and the voltage is ramped up to 50kV (gradient 50MV/m); Phase II – gradient is kept constant (50MV/m) but the gap is increased up to 7mm (350kV). In phase III voltage is kept constant (350kV) and the gap is closed down gradually until a breakdown occurs.

### III. ELECTRODE MATERIALS AND SURFACE TREATMENTS FOR HIGH GRADIENT ELECTRODES IN VACUUM

#### A. Bare metal electrodes

Surface finish of metal electrodes plays the major role in breakdown strength. From extensive comparison, it was found that hand polishing was necessary for the best performance. Fig. 4 summarizes the achieved breakdown strength for different polished metals. Gray bars indicate materials' tensile strength.

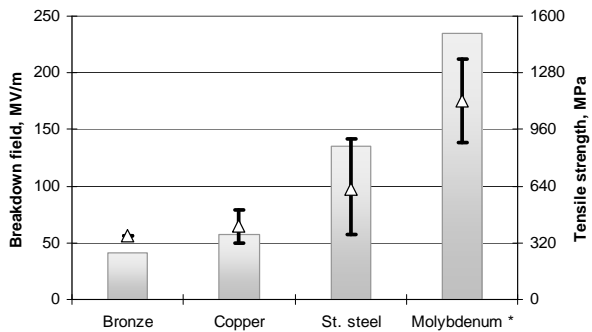

Figure 4.  Polished metal electrodes comparison.
The 2μm molybdenum was sputtered on polished stainless steel.

Unlike the other metal samples, molybdenum was sputtered on polished stainless steel surface. It is interesting to notice the correlation between breakdown strength and tensile strength of the material. (In Fig. 4, for sputtered molybdenum, the bulk tensile strength value is indicated.) The triangles indicate the average breakdown value and the error bars show the full span of the results. There was no significant difference in achieved breakdown strength for in-house and externally hand polished electrodes. Fig. 5 illustrates the typical surface quality of hand polished stainless steel electrode with roughness figure Ra = 13nm.

In general, maximum achieved breakdown gradients for bare metal surfaces did not exceed 150MV/m

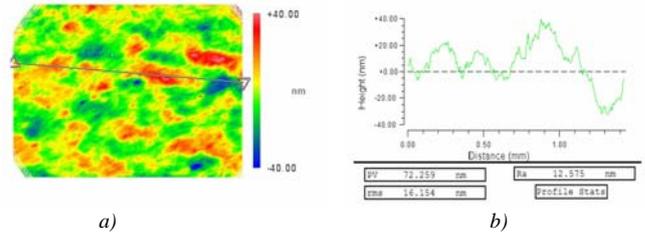

Figure 5.  *a)* 2D vertical surface mapping and *b)* line vertical profile of polished stainless steel electrode.

#### B. DLC coated electrodes

Due to their unique mechanical and electrical properties DLC coatings were an attractive candidate for further breakdown strength improvement of metal electrodes. Two main factors are expected to increase the breakdown strength. The first one is the thin and smooth dielectric layer (relative electrical permittivity $\varepsilon_r > 1$) will reduce metal surface electric field with a factor equal to $\varepsilon_r$, reducing the overall metal surface field. The reduced surface field should reduce electron field emission that is believed to play a major role in initiating a breakdown process. The second factor is materials with higher tensile strength are expected to have higher breakdown strength. Some hydrogenated amorphous DLC films have more than one order of magnitude higher tensile strength compared to steel.

Using Plasma Assisted Chemical Vapor Deposition (PACVD) process it is possible to deposit hydrogenated amorphous DLC (a-C:H) with custom properties (coating thickness, hardness and conductivity) on virtually any type of metal surface (www.bekaert.com). The first breakdown tests with DLC coated electrodes gave unexpectedly good results. Later, several different coatings, from different suppliers were compared.

The influence of five coating parameters on breakdown strength were explored: DLC layer thickness, DLC layer type (electrical resistivity), base metal, base metal surface roughness and coating process from different suppliers.

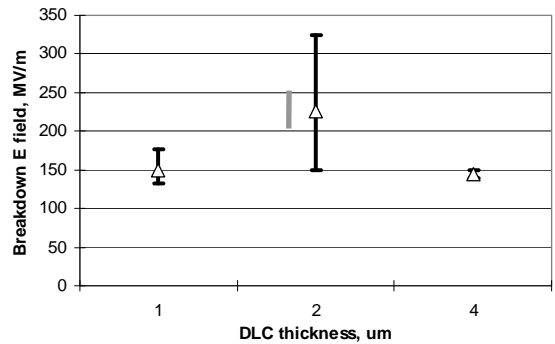

Figure 6.  DLC layer thichness comparison (a-C:H, Bekaert). Base metals: bronze, copper and stainless steel.

Fig. 6. summarizes the results for three coating thicknesses (coating type: a-C:H, Bekaert). The error bars represent the full span of measured results. For 1μm and 4μm, the number of samples is small. The large spread for 2μm is dominated by the bronze samples (gray bar represents stainless steel results only). Fig. 6. is a compilation of results with different base metals: bronze, copper and stainless steel. Due to the limited number of samples Fig. 6. should be used as a rough comparison only. Nevertheless, the opinion is that 2μm thick coating gives the best results because the coating process has been optimized for it.

Fig. 7. shows the results for three different coating types (Bekaert) and their micro hardness (gray bars). The base metal is stainless steel. There is some correlation between the breakdown performance and micro hardness.

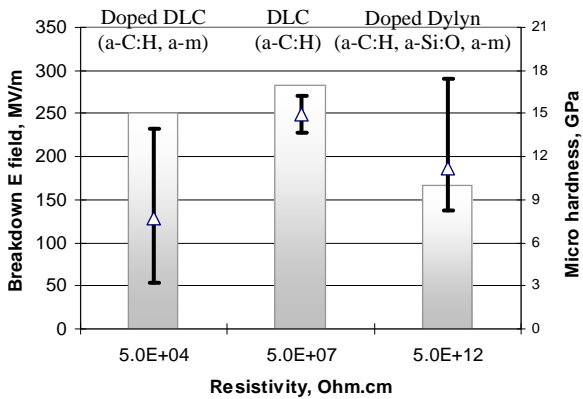

Figure 7. Coating types comparison. 2μm DLC deposited on polished stainless steel (Bekaert).

Fig. 8. compares the results for three different base metals with 2μm DLC from Bekaert. The triangles represent the average value and the error bars give the full span of the measured results. Some DLC coatings had macroscopic defects. In Fig. 8., the two low values for bronze are probably related to such defects. Copper values will be higher than shown because some of the samples were saved for future electron beam experiments and were not tested until breakdown.

Rougher base metal surface may give lower breakdown value – only one sample was tested (stainless steel, Bekaert, 2μm DLC, sand paper finish, not polished, breakdown field 122MV/m).

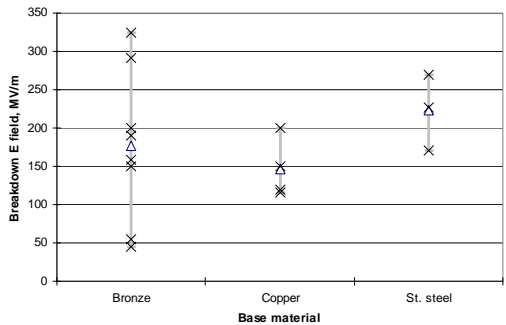

Figure 8. Comparison of base metals (Bekaert, 2μm DLC).

DLC coating from different suppliers were compared as well. All of them used PACVD deposition process except Fraunhofer Institute which used Ion Beam Sputter Deposition (IBSD). Fig. 9. compares the results for 2μm DLC deposited by different companies on polished stainless steel.

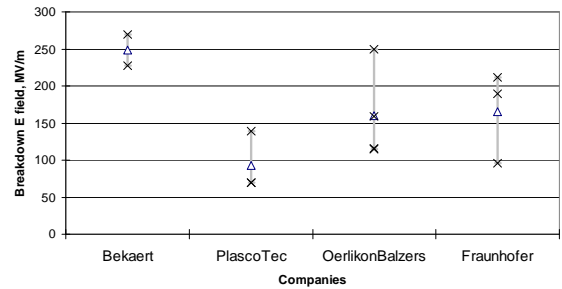

Figure 9. 2μm DLC coating from different suppliers on polished stainless steel.

IV. DLC PHOTOEMISSION

The exact DLC photoemission process is not yet very well understood [2]. Moreover, doped DLC films (such as Diamond Like Nanocomposite) have different and not well defined electrical and mechanical properties due to their dependency on sp2/sp3 bonding ratio, doping levels and coating process parameters [3]. Fig. 10. illustrates schematically the DLC layer structure from Bekaert and two possible photo emission mechanisms.

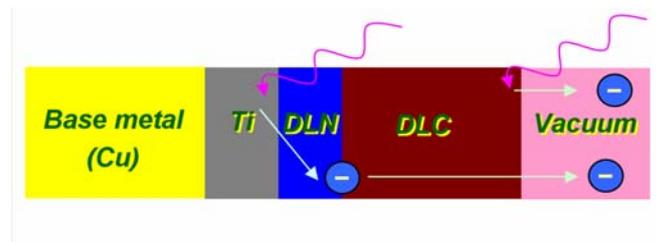

Figure 10. Typical DLC structure (Bekaert) and the two possible photo emission mechanisms.

In the test stand, DLC coated electrodes were illuminated with UV laser (262nm) and electron bunches were emitted in large surface gradient range (25 to 170MV/m). Fig. 11. compares measured Quantum Efficiency (QE) with theoretically calculated one based on single barrier photoemission model [4].

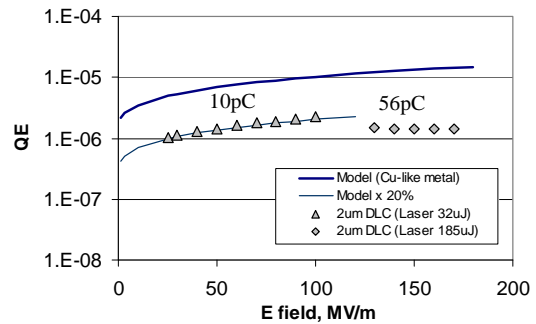

Figure 11. Photo emission from DLC coated cathode.

This model assumes photo electron injection from copper-like metal (work function $\phi$ = 4.6eV) to DLC conduction band and electrons' transport through the DLC layer. In the model the gradient at metal-DLC interface is reduced 4 times taking in account DLC dielectric constant and 25% light transmission through the DLC layer. At low charge the model predicts 5 times higher quantum efficiency.

## V. DLC Hollow Cathode

The high breakdown strength of DLC coated electrodes made it possible to develop so called "hollow" cathode geometry for testing different photo-emitting materials and FEAs. The hollow cathode lip cover the edge of the sample and makes electrical contact to the sample's front surface. In addition, electric field lines in proximity of the emission surface are deformed due to concave electrode profile, providing electrostatic electron beam focusing. This is valuable because electrons have small kinetic energy and the beam is prone to space charge degradation without additional focusing. Fig. 12. shows a cross section of the hollow cathode.

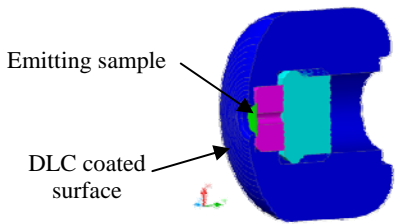

Figure 12. Hollow cathode cross section.

## VI. Electron Emitters

### A. Metallic samples

Different metals were tested for efficient photoemission. The samples were hand polished in air. A few minutes before installation, the last polishing step was repeated. In this way metal surface exposure to air was reduced. Samples' surface is cleaned using dry snow blasting and no further surface preparation is applied. Figure 13. summarizes photo emission QE results for different metals.

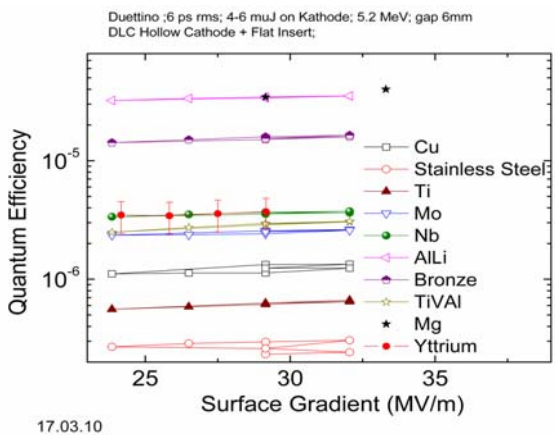

Figure 13. Photo emission QE of different metal samples.

### B. Field Emitting Arrays

The hollow cathode geometry also made it possible to test fast, electrically gated FEAs in high gradient environment. Electron bunches of 4ns (FWHM) duration with up to 300keV kinetic energy were produced. Maximum extracted charge was more than 10pC and maximum FEA surface gradient was 30MV/m. Fig. 14. shows the structure of the emitted electron beam from a 2mm diameter FEA. Emission homogeneity needs to be improved.

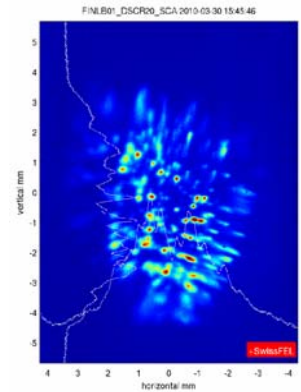

Figure 14. FEA electron beam image.

## VII. Conclusion

It was found that hydrogenated amorphous DLC (a-C:H) coating of metal electrodes improves significantly their vacuum breakdown strength for damped oscillatory electric pulses. Using DLC coated electrodes, more than 250MV/m accelerating electric field was achieved for electrodes distance of ~1mm. DLC coated metal surfaces showed stable photo emission and up to 60pC electron bunches (@350keV) were emitted in accelerating gradient up to 150MV/m.

Hollow geometry cathode with DLC coating was developed in order to integrate photoemission samples and FEAs into the cathode, also giving electrostatic electron beam focusing close to emitting surface. QE of different metal photo cathodes was measured.

Fast electrically gated FEAs were also integrated into the hollow geometry cathode. Maximum electric field on the FEA surface was 30MV/m and maximum beam energy was 300keV; higher values were possible. Maximum extracted charge was more than 10pC (4ns FWHM) with the FEA surface gradient of 9MV/m and beam energy of 250keV. It was proven that an FEA could operate in a HG environment but emitted beam homogeneity needs improvement.